\documentclass{trbunofficial}
\usepackage{graphicx}
\usepackage{algorithm, algpseudocode} 
\usepackage{newtxtext}
\usepackage{newtxmath}
\usepackage{enumitem}
\usepackage[hidelinks]{hyperref}
\AuthorHeaders{Tian, Feng, Zhou, Anupriya, Quddus, Demiris, and Angeloudis}
\title{The Impact of Building-Induced Visibility Restrictions on Intersection Accidents}
\author{%
  \textbf{Hanlin Tian, Corresponding Author}\\
  Department of Civil and Environmental Engineering\\
Imperial College London, London, UK, SW7 2BU\\
  Email: h.tian22@imperial.ac.uk\\
  \hfill\break%
  \textbf{Yuxiang Feng}\\
    Department of Civil and Environmental Engineering\\
Imperial College London, London, UK, SW7 2BU\\
  Email: y.feng19@imperial.ac.uk\\
  \hfill\break%
  \textbf{Wei Zhou}\\
    Department of Civil and Environmental Engineering\\
Imperial College London, London, UK, SW7 2BU\\
  Email: wei.zhou22@imperial.ac.uk\\
  \hfill\break%
  \textbf{Anupriya}\\
    Department of Civil and Environmental Engineering\\
Imperial College London, London, UK, SW7 2BU\\
  Email: anupriya15@imperial.ac.uk\\
  \hfill\break%
  \textbf{Mohammed Quddus}\\
    Department of Civil and Environmental Engineering\\
Imperial College London, London, UK, SW7 2BU\\
  Email: m.quddus@imperial.ac.uk\\  
  \hfill\break%
  \textbf{Yiannis Demiris}\\
    Department of Electrical and Electronic Engineering\\
Imperial College London, London, UK, SW7 2BU\\
  Email: y.demiris@imperial.ac.uk\\
  \hfill\break%
  \textbf{Panagiotis Angeloudis}\\
    Department of Civil and Environmental Engineering\\
Imperial College London, London, UK, SW7 2BU\\
  Email: p.angeloudis@imperial.ac.uk
}

\begin{document}
\maketitle
\section{Abstract}
Traffic accidents, especially at intersections, are a major road safety concern. Previous research has extensively studied intersection-related accidents, but the effect of building-induced visibility restrictions at intersections on accident rates has been under-explored, particularly in urban contexts.

Using OpenStreetMap data, the UK's geographic and accident datasets, and the UK Traffic Count Dataset, we formulated a novel approach to estimate accident risk at intersections. This method factors in the area visible to drivers, accounting for views blocked by buildings - a distinctive aspect in traffic accident analysis.

Our findings reveal a notable correlation between the road visible percentage and accident frequency. In the model, the coefficient for "road visible percentage" is 1.7450, implying a strong positive relationship. Incorporating this visibility factor enhances the model's explanatory power, with increased R-square values and reduced AIC and BIC, indicating a better data fit.

This study underscores the essential role of architectural layouts in road safety and suggests that urban planning strategies should consider building-induced visibility restrictions. Such consideration could be an effective approach to mitigate accident rates at intersections. This research opens up new avenues for innovative, data-driven urban planning and traffic management strategies, highlighting the importance of visibility enhancements for safer roads.

\hfill\break%
\noindent\textit{Keywords}: Road Safety, Intersection Visibility, Accident Prediction, Geographic Information Systems (GIS)

\newpage

\section{Introduction}

Road traffic accidents remain a persistent global public health concern, leading to severe injuries, fatalities, and substantial economic impact annually \cite{WHO2020}. As suggested \citet{who2022}, every year there is about 1.3 million people could not survive the catastrophic traffic accidents, and a further 20-50 million people suffer from non-fatal injuries. In addition to the human casualties, road traffic accidents result in significant economic losses for most countries, representing approximately 3\% of their Gross Domestic Product (GDP) \cite{who2022}. These accidents are influenced by a multitude of factors, including road conditions, weather, driver behavior, and vehicle characteristics \cite{lee2014analysis, tomoda2022analysis, huang2008severity, useche2018infrastructural, elvik2009handbook}. However, one element that remains inadequately explored is the role of visibility at intersections, particularly within the dense urban fabric of city landscapes. Our study aims to bridge this understanding gap by conducting a GIS-based quantitative examination of the correlation between road visibility and accident occurrence rates in the city of London. We further extended our research scope to the city of Manchester to validate the generalization of our model and research findings.

Intersections are known hotspots for traffic accidents due to the highly frequent interactions among different types of road users \cite{laureshyn2010evaluation}. The National Highway Traffic Safety Administration (NHTSA) reports that over 80\% of pedestrian accidents relate to intersections\cite{nhtsa}. The risk of collisions at intersections is approximately 335 times higher than at non-intersections, primarily due to obstructed sightlines. \citet{kumar2016data} employed data mining to classify accident-prone locations based on frequency. Yet, in the effort to enhance intersection safety, one potential risk factor has been largely overlooked—the potential impact of obstructed driver visibility due to surrounding buildings \cite{turner2006predicting}. Buildings adjacent to intersections can obscure drivers' views of incoming traffic, pedestrians, or road signs, thereby increasing the likelihood of accidents. This study innovatively incorporates the key factor of buildings into the calculation of intersection visibility through GIS spatial image analysis. The resulting visibility index is then integrated into the intersection risk assessment, enabling a more comprehensive understanding of the accident-proneness of intersections based on spatial analysis.

Our study highlights the significant role of intersection visibility in road traffic accidents. The insights gained can inform urban planners, traffic engineers, and policymakers in designing safer road networks and formulating effective traffic strategies. Ultimately, our findings contribute to efforts aimed at reducing the frequency of road traffic accidents, improving overall road safety in urban environments \cite{elvik2009handbook}.

Our study makes the following contributions to the field of traffic safety and urban planning:

\begin{itemize}
\item Our approach is the first to consider the driver's field of view as a crucial element in intersection safety evaluation. This innovative methodology provides fresh insights and potential strategies for enhancing traffic safety.
\item We introduce a real-time method for estimating intersection risk, leveraging live traffic and environmental data. This method enables dynamic responses, a critical aspect of improving traffic management efficiency.
\item Our model, tested extensively with the London map and the UK accident dataset, showed a substantial decrease in AIC and BIC by 322.8 and 317.93 units respectively when the 'view percentage' was included, strongly suggesting its significance as a predictor of accidents. 
\end{itemize}

Our paper is organized as follows:

\begin{enumerate}

\item \textbf{Related Work}
-This section reviews previous studies related to the research topic. It provides a critical evaluation of the existing literature, identifying gaps or inconsistencies in the current knowledge, and outlining how the present study aims to address these gaps.
\item \textbf{Data Collection and Preprocessing}
- This section details the data sources used in the study and how the data was gathered. It also describes the preprocessing steps taken to prepare the data for analysis.

\item \textbf{Methodology}
- This section explains the methods used in the study, including the calculation of view percentages, the categorization of road sections, and the intersection risk modeling process.

\item \textbf{Results and Discussion}
- This section presents the results obtained from the analysis, including the model performances and the relationships identified. The findings are discussed in the context of the research questions and the existing literature.

\item \textbf{Conclusions and Future Work}
- This section provides a summary of the research and its key findings. It outlines the implications of the results, draws conclusions, and suggests potential areas for future research.
\end{enumerate}

\section{Related Work}

\subsection{Intersection as Critical Zones for Road Accidents}

Intersections, with their heightened number of conflict points and requirements for simultaneous interactions among multiple road agents, are well recognized as significant hotspots for road safety concerns. The likelihood of collisions at intersections is reported to be approximately 335 times higher than at other road segments, predominantly caused by vehicles turning with obstructed sightlines \cite{congress2021identifying}. Furthermore, the National Highway Traffic Safety Administration reports that over 80\% of pedestrian accidents can be categorized into specific types, with a substantial portion being intersection-related \cite{nhtsa}.

Several studies have investigated this intersection-accident nexus. A study by \cite{al2003analysis} analyzed the triggers of intersection-related accidents, identifying key traffic safety issues. However, this study did not dissect the unique characteristics of intersection and non-intersection-related accidents and lacked an in-depth exploration of geographic features and traffic characteristics. Similarly, \citet{kumar2016data} used data mining techniques to classify accident locations based on accident frequency, but their analysis was limited by the lack of additional data, such as vehicle speed, weather information, and road conditions.

\subsection{Exploration into Intersection Accident Factors}

The complexity of intersection-related traffic accidents has prompted a multitude of research efforts aimed at understanding the contributing factors. These factors can be broadly divided into two major categories: human factors and environmental factors.

Investigations into human factors have largely centered around the behaviors of drivers and pedestrians. Studies, such as those by \citet{lee2014analysis} and \citet{tomoda2022analysis}, have probed the correlation between population size, non-motorized transport users, children's behaviors at intersections, and accident rates. However, these studies face limitations due to the reliance on self-reported data and the lack of geographical and cultural standardization. While qualitative data offer insights into individual experiences, perceptions, and subjective interpretations, the integration of quantitative metrics, such as accident rates, can provide a more comprehensive and measurable understanding of human factors contributing to intersection-related accidents.

On the other hand, research into environmental factors in intersection-related accidents employs various quantitative methodologies and models. These studies consider variables like accident severity, road type, external conditions, intersection attributes, time factors, infrastructure elements, and vehicle factors \cite{dixon2015improved, eboli2020factors, huang2008severity, oh2004development, mussone2017analysis, gong2020application, chen2012analysis, park2016random, lam2004environmental}. However, many of these studies overlook less tangible factors such as geographical characteristics of accidents and other potential contributors that are difficult to directly observe or record.

\subsection{Intersection Visibility and Road Accidents}

Intersection visibility, encompassing the driver's visual field and sight distance, plays a pivotal role in road safety \cite{todd2017quantitative, bauer1996statistical}. Sight distance, or the length of road visible to a driver, is particularly crucial at intersections to ensure safe maneuvering. However, various factors such as road curvature or roadside obstructions can compromise this visibility. In response to these challenges, recent studies have employed high-resolution lidar data, simulations, and Digital Elevation Models (DEM) to explore these elements and assess their impact on sight distance \cite{chen2012analysis}.

Despite these advancements, there exists a notable gap in comprehensive research linking sight distances, fields of view, and intersection-related risks. Our research aims to bridge this gap by employing information. By doing so, we hope to offer a more holistic investigation of intersection visibility and its implications for road safety \cite{mussone1999analysis}.

\subsection{The Role of Geographic Information Systems (GIS) in Accident Analysis}

Intersection accident analyses often use macro-factor datasets from broad geographical areas. They typically account for accident counts and severity over specific periods. Many past studies target specific regions, but a few have analyzed accident datasets at a macro level \cite{ma2014quasi, hu2020investigation, chen2019multinomial, gomes2013influence}. However, this might lead to spatial autocorrelation issues, challenging the independent observation regions assumption \cite{dale2002spatial}. Thus, spatial considerations are vital for these investigations.

GIS tools, essential for geospatial data analysis, are becoming more prevalent in road safety studies. Existing literature mainly discusses two GIS applications: identifying accident hotspots and geocoding accidents for spatial categorization and analysis. In hotspot identification, techniques include kernel density estimation, nearest neighbor distances, and spatial indices like Moran's I. For example, \cite{al2021mapping} used the SANET toolkit to assess hotspot locations, while \citet{cheng2018traffic} applied the Getis-Ord Gi* technique for spatial autocorrelation. The second category, geocoding, facilitates statistical evaluations. \citet{ma2014quasi} visualized geocoded road and accident datasets, linking accidents to road segments. Similarly, \cite{loo2006validating} integrated the Traffic Accident Data System with road networks to position accident data accurately. \cite{noland2004spatially} spatially organized England's road casualty data, connecting it to land-use types in electoral districts.

In conclusion, while previous studies have contributed significantly to understanding the factors influencing intersection-related accidents, a gap persists in the literature with respect to the incorporation of spatial and visibility factors at the micro-level. Our research aims to bridge this gap by integrating human factors, environmental factors, and GIS capabilities, with particular emphasis on the quantification of visibility at intersections. By doing so, we introduce a novel perspective to the study of intersection-related accidents. This approach allows us to capture the subtle yet significant influence of intersection visibility on accident occurrence, thereby contributing to a more nuanced understanding that could inform more effective road safety measures.

\section{Methodology}

\begin{figure}[ht]
\centering
\includegraphics[width=0.6\textwidth]{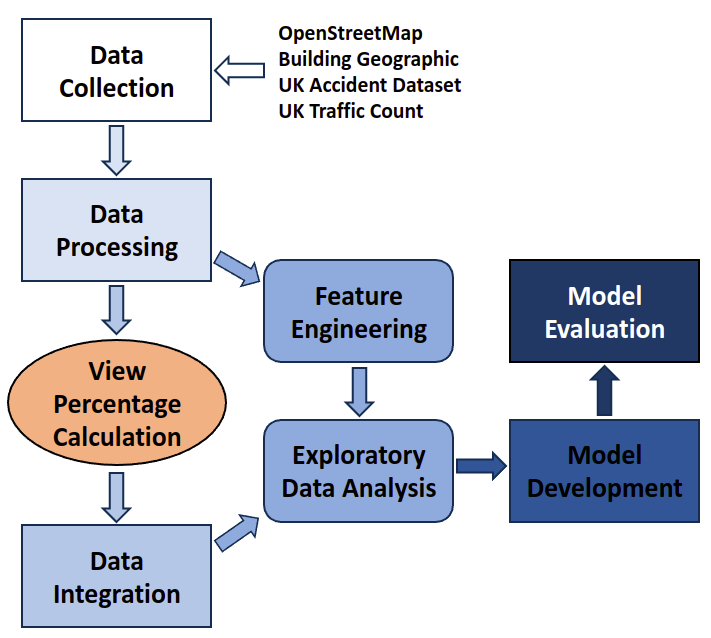}
\caption{Flowchart of methodology}
\label{fig:flowchart}
\end{figure}

In our study, we follow a systematic methodology to explore the factors influencing road accidents, as illustrated in Figure \ref{fig:flowchart}. We begin with data collection, assembling key datasets from diverse sources including OpenStreetMap for intersection details, the UK Accident Dataset for historical accident records, the UK Traffic Count for traffic volume statistics, and the Building Geographic Dataset for information about the built environment surrounding each intersection.

After collecting data, we proceed to data processing. This involves geocoding each accident to the nearest intersection, cleaning the data, and imputing missing values. Feature engineering is conducted based on preliminary data analysis.

Once processed, we merge the cleaned datasets into a unified one for further analysis, which underpins our comprehensive exploration and modeling.

We then explore the data to grasp its structure and discern patterns. A crucial aspect of this exploration phase involves computing the unobstructed view percentage at each intersection. Our hypothesis posits that this factor may exert a substantial negative influence on the occurrence of accidents.

\begin{table}[ht]
\centering
\caption{Datasets and Important Element}
\begin{tabular}{|l|l|}
\hline
\textbf{Dataset} & \textbf{Important Elements} \\
\hline
OpenStreetMap Data & Intersection details, Road types, Traffic controls \\
\hline
UK Accident Dataset & Accident location, Time, Vehicles involved \\
\hline
UK Traffic Count Dataset & Average daily traffic volume \\
\hline
Building Geographic Dataset & Building coordinates, Physical dimensions\\
\hline
\end{tabular}
\label{table:datasets}
\end{table}

Finally, we build and train our risk models tailored to intersections. We utilize the Generalized Linear Model (GLM) framework and assume a Poisson distribution for accident counts, which is widely used for modeling count data. The primary objective of these models is to discern and quantify the correlation between diverse intersection characteristics and accident risk. By undertaking this rigorous analysis, we strive to extract actionable insights that can serve as a foundation for informing future traffic safety measures and policy decisions, ultimately contributing to the improvement of intersection safety and accident prevention efforts.

\subsection{Data Source}
In our study, we leveraged three primary data sources: OpenStreetMap \cite{haklay2008openstreetmap} for road network information and intersection details, the UK Road Accident and Safety dataset\cite{ukaccidentdata} for historical accident records, and the UK Traffic Count\cite{uktrafficcount} for traffic volume statistics based on coordinates. Additionally, we utilized the architectural geographic dataset to construct the physical dimensions of buildings around the road network based on coordinates. By geographically encoding all three databases into a single Geographic Information System, we can more comprehensively and accurately combine space to analyze the factors affecting road accidents. The details and the important elements from each dataset used in our study are summarized in Table \ref{table:datasets}.

\subsubsection{OpenStreetMap Data}

OpenStreetMap (OSM)
OSM is a collaborative project aimed at creating a free and editable map of the world \cite{haklay2008openstreetmap}. It offers the capability to map and describe real-world features, ranging from administrative boundaries and topographical characteristics to amenities and street infrastructure. Within this, road networks and buildings stand out as significant landmark elements due to their pivotal roles in urban transportation and architectural structures. From OSM, we extracted geographical and infrastructural details about intersections, including road types and traffic controls. While OSM data is created by a multitude of contributors leading to occasional inconsistencies, it remains a highly valuable resource due to its detail, up-to-dateness, and free availability.  In addition to manual mapping by contributors, some data has been cross-verified with authoritative sources\cite{witt2021analysing}. Due to the collaborative nature of OSM, it encompasses information on slums and informal residential areas more extensively than other map data\cite{panek2015community}, providing deeper practical relevance to our research.

\subsubsection{Building Geographic Dataset}

The Building Geographic Dataset, derived from OpenStreetMap, serves as an instrumental tool in our research, offering exhaustive geospatial information on the constructed landscape within London. In the architectural geographic dataset, the mapping function relies on the topological configuration of two primary components: vertices and closed paths. Vertices serve as the fundamental mapping elements, denoting their geographical coordinates in terms of latitude and longitude. Closed paths, formed by the continuous connection of vertices, create polygonal structures commonly used to represent the external shape of buildings\cite{biljecki2023quality}. The dataset also includes comprehensive details like the geographical coordinates of buildings, their physical dimensions (height, footprint area, etc.), and their intended usage (residential, commercial, etc.). 

These attributes contribute to the understanding of the built environment around each intersection. The size and placement of buildings can substantially affect the visibility at an intersection, influencing drivers' ability to perceive potential hazards. 

\subsubsection{UK Accident Dataset}

The UK Accident Dataset\cite{ukaccidentdata}, maintained by the UK Department for Transport, provided us with comprehensive historical records of road accidents reported to the police. This dataset included details such as the accident's latitude and longitude location, vehicles involved, casualties, and road and weather conditions at the time of the accident. This dataset allowed us to locate accident-prone intersections and provided a wide range of variables for our analysis.

\subsubsection{UK Traffic Count: Annual Average Daily Flow (AADF)}

The Annual Average Daily Flow (AADF) data \cite{uktrafficcount}, which gives an estimate of average daily traffic volume on a particular section of road, was another crucial component of our study. Higher AADF values indicate greater traffic volume, potentially increasing accident likelihood \cite{glushkov2021analysis}. However, the relationship between traffic volume and accidents isn't direct and needs to be considered alongside other variables like road type and speed limit.

\subsection{Data Gathering and Preparation}

\subsubsection{Data Collection}

The road accident data and the UK Traffic Count Dataset are sourced directly from the UK government's Department for Transport. These datasets provide essential details about the accident's location, time, severity, the types of vehicles involved, and the average daily traffic volume at various locations across the UK.

For intersection-related geographic data, we utilize the OSMnx library. It facilitates the direct download and modeling of transport network data from OpenStreetMap, providing detailed information about the physical attributes of an intersection.

\subsubsection{Data Processing}

In the feature engineering step, we created new, more informative variables from our original features. One such feature was the count of accidents within a certain boundary around each intersection. To achieve this, we created a buffer around the center point of each intersection and counted the number of accidents that fell within this area.

In previous research, the use of buffer zones sometimes resulted in a vast circular boundary area\cite{loo2006validating}. When overlaps occurred, certain accident points might fall within multiple buffer zones, leading to issues of double-counting and ambiguous delineation. As a result, we experimented with various buffer radii to ensure that all accidents occurring near intersections were encompassed, while avoiding the inclusion of unrelated accidents in the analysis. After several iterations, we determined that a radius of 0.0003 units (approximately equal to 33.3 meters) was the most effective in capturing accidents associated with each intersection, striking a balance between accuracy and specificity in our accident counting.

In our data processing, we tailored our focus to include only specific road and intersection types and more recent years. We limited our analysis to primary road intersections, excluding minor roads and non-major junctions like roundabouts and slip roads. This focus allowed us to hone in on intersections that are most relevant to our study. Additionally, to account for potential changes in the road network over time due to urban development, road improvements, and evolving traffic patterns, we constrained our analysis to data post-2010. This ensured our findings were based on the most current and accurate representation of London's road network.

We also engineered a feature to represent the estimated road width. To achieve this goal, we created a rectangular custom buffer zone around each road segment represented by a line segment. The width of this buffer was determined based on the standard road width in London, which was converted to geographical units (degrees). Specifically, we used a conversion factor of 0.000133 units per meter (based on the latitude of London) to translate the standard road width of 14.8 meters into geographical units. This estimated road width, expressed in geographical units, was then used in our analyses.

\begin{figure}[ht]
\centering
\includegraphics[width=0.8\textwidth]{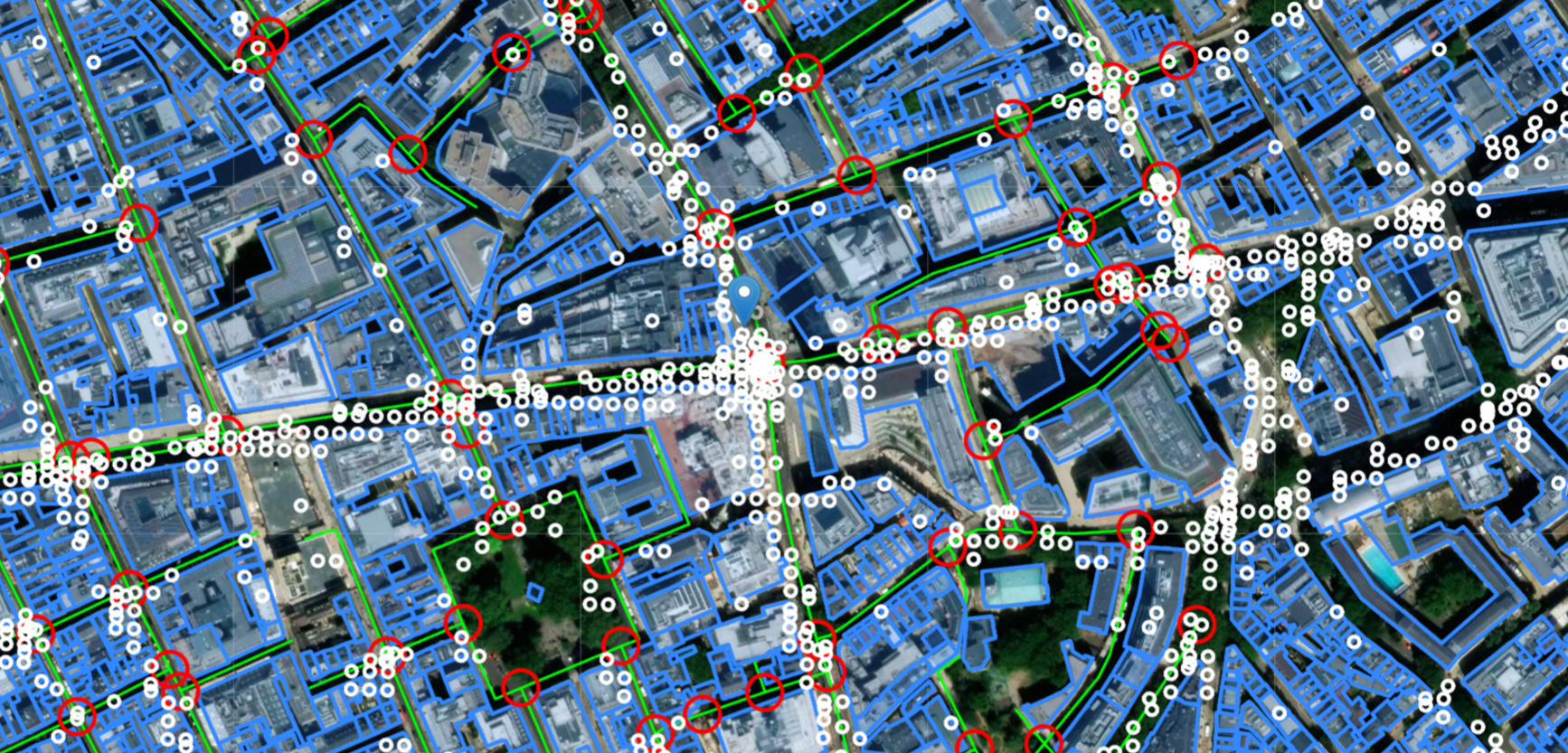}
\caption{The plot of the map using Folium. The blue polygons represent buildings, red circle represents each intersection, and each white circle is an accident.}
\label{fig:map}
\end{figure}

\subsubsection{Data Integration}

After cleaning and processing our individual datasets, we then integrated the road accident data and intersection data into one unified dataset. To facilitate this, we created a dictionary where the key was the OpenStreetMap ID (osmid) of each intersection, and the value was a comprehensive record including all corresponding edges, the number of accidents, and other relevant parameters such as road type and traffic volume.

This data structure allowed us to efficiently link accident data with the corresponding intersection characteristics. The resulting integrated dataset provided a rich source of information for our subsequent analyses.

\subsubsection{Data Exploration}

In the data exploration phase, we visualized the spatial distribution of accidents using the Folium library in Python. The generated map, as illustrated in Figure \ref{fig:map}, marks accident locations with white circles and intersections with red circles. The surrounding buildings are represented by blue polygons. To enhance the visualization, we incorporated the 'Mapbox Satellite' as the base map, which aids in identifying potential accident hotspots and deriving hypotheses for further analysis.

\subsection{View Percentage Calculation}
\begin{figure}[ht]
\centering
\includegraphics[width=0.7\textwidth]{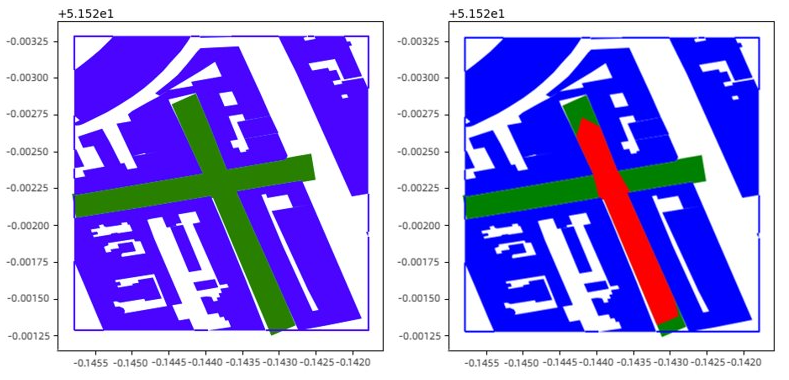}
\caption{Buffued lane display and view area display}
\label{fig:buffed}
\end{figure}

Our study notably highlights the role of visibility in road accidents. We developed a methodology to calculate the 'view percentage' at each intersection, a measure representing the proportion of unobstructed view from a given point on a road section. We hypothesized that a lower view percentage, indicating limited visibility, might increase the likelihood of accidents.

The view percentage calculation was guided by established standards for assessing visibility at intersections \cite{manary1998development, todd2017quantitative}. These standards suggest a 80-degree field of view as a reasonable approximation for a driver's effective field of view at an intersection. This angle forms the basis for our visibility calculations.

Our methodology involves computing the angles of view blocked by buildings around an intersection. The function 'CalculateAngles' in our algorithm computes the start and end angles of the view blocked by a given building relative to an intersection. By aggregating the blocked views from all surrounding buildings, we estimate the total blocked view. This process is illustrated in Figure \ref{fig:buffed}, where the red areas represent the blocked view caused by buildings.

The view percentage is then calculated as the ratio of unblocked view to the total view. The total view is represented as a full circle, with an arc of 2$\pi$ radians.

The pseudocode for the view percentage calculation algorithm is as follows:

\begin{algorithm}
\caption{\textbf{View Percentage Calculation}}
\begin{algorithmic}[1]
\Procedure{CalculateViewPercentage}{$intersection, buildings$}
\State $total\_angle \gets 2*\pi$
\State $blocked\_angle \gets 0$
\For{each $building$ in $buildings$}
\State $angle\_start, angle\_end \gets$ \Call{CalculateAngles}{$intersection, building$}
\State $blocked\_angle \gets blocked\_angle + (angle\_end - angle\_start)$
\EndFor
\State $view\_percentage \gets 1 - \frac{blocked\_angle}{total\_angle}$
\State \Return $view\_percentage$
\EndProcedure
\end{algorithmic}
\end{algorithm}

The implementation of this algorithm in a real-world dataset involves handling additional complexities. For instance, we encounter different geometrical objects representing road sections and buildings. Roads could be represented as 'LineString' or 'MultiLineString', requiring us to handle each 'LineString' in a 'MultiLineString' separately. We also use spatial operations from the geometric library such as 'intersection' to compute the intersection between the road and buildings, and 'buffer' to create a buffer around the road section.

Additionally, the calculation of the field of view percentage involves interpolation at points along each road segment. To best simulate the genuine change in viewpoint as an agent moves around intersections, we established multiple interpolation points on each segment to create new viewpoints. Each interpolation point serves as the starting point for a sectorial view, producing a consistent 80-degree field of view, with considerations for obstructions caused by buildings. This is achieved by the following method: for each interpolation point, a loop is set up which rotates around the point to produce 80 rays, ranging from -40 degrees to 40 degrees, with a 1-degree rotation each time. Subsequently, the multiple intersections of the rotated line segments with the building outlines are calculated, with the nearest intersection to the interpolation point (the starting point of the view) chosen to form line segments. Constructing a polygon with these line segments results in the field of view region considering building obstructions. After calculating the fields of view for all intersections, various interpolation distances were tested. We chose an interpolation distance that strikes a balance between computational efficiency and the granularity of visibility calculations. The interpolation process will generate a set of points along each segment and calculate the view percentage for each point.

\begin{figure}[ht]
\centering
\includegraphics[width=0.9\textwidth]{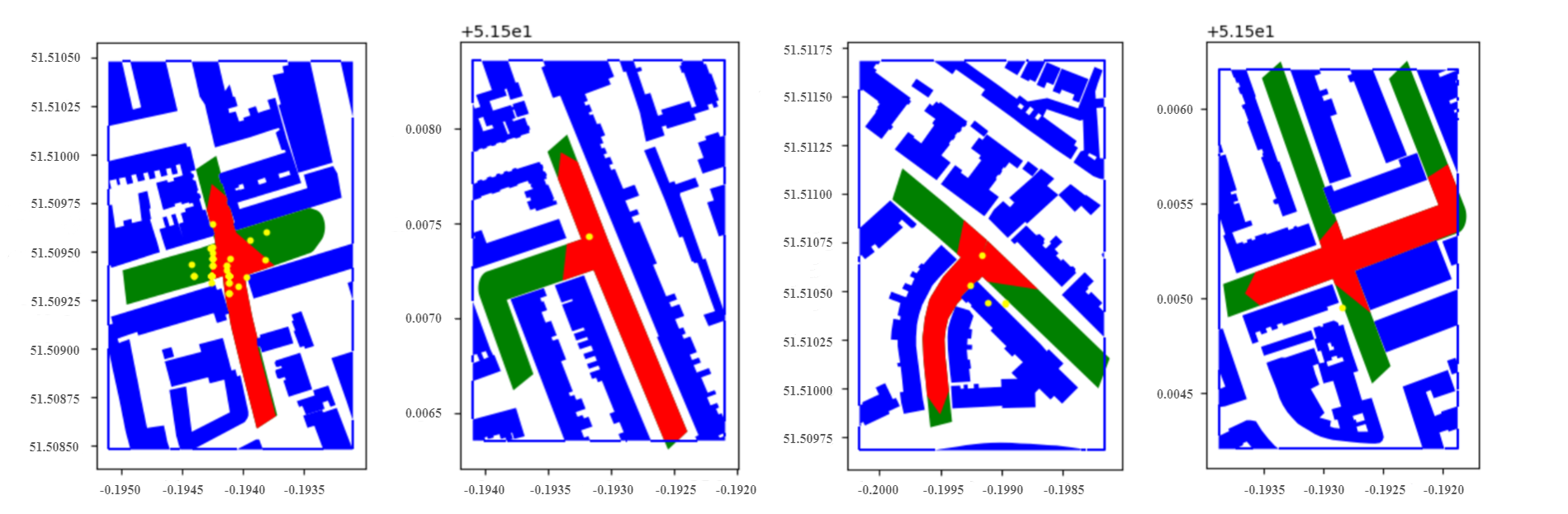}
\caption{Result plot of occluded view area : visible area (red) vs. road area (green)}
\label{fig:occlusion1}
\end{figure}

Figure \ref{fig:occlusion1} provides a visual illustration of the view percentage calculation. It depicts an intersection, the surrounding buildings, and the blocked and unblocked views from a point on the road section. This graphical representation gives us a visual sense of how the view percentage is determined and how it relates to the built environment around the intersection.

By integrating the view percentage into our dataset, we aim to capture the impact of road visibility on accident rates, providing a more comprehensive understanding of the factors contributing to road accidents.

\subsection{Categorizing Road Sections and Traffic Flow Assignment}

In our study, road sections are categorized into 'primary' and 'secondary' types based on the nature of the intersecting roads at an intersection. If any road intersecting at an intersection is classified as 'primary', we categorize the intersection and its associated road section as 'primary'. Conversely, if an intersection does not connect to a 'primary' road, we designate the intersection and its associated road section as 'secondary'. This classification scheme is grounded in the assumption that the type of a road section is largely influenced by the most significant road type it interacts with.

In addition to road type, we also assign traffic volume to each intersection. To achieve this, we use a spatial index to locate the nearest traffic count point for each intersection and then assign the traffic flow of that point to the intersection. This process allows us to incorporate the influence of traffic volume, a critical factor contributing to road accidents, into our analysis.

\subsection{Intersection Risk Modeling}

In the data analysis stage, the primary objective is to uncover patterns and relationships in the dataset by employing a combination of statistical and machine learning methods \cite{hastie2009elements}. The analysis encompasses an initial exploratory data examination, enabling us to gain fundamental insights into the underlying characteristics of the data. Subsequently, we delve into more sophisticated analysis techniques such as regression analysis and classification. These methods allow us to identify influential factors associated with road accidents at intersections, ultimately contributing to a comprehensive understanding of primary factors affecting intersection safety.

Our analysis places special emphasis on the role and characteristics of intersections, investigating the propensity of certain types of intersections toward accidents. We delve into the correlation between intersection features and accidents, exploring if and how specific combinations of features could particularly heighten risk. The objective is to identify critical intersection characteristics that significantly influence accident occurrence, thereby informing targeted interventions for road safety improvement.

Two models based on the Generalized Linear Model (GLM) framework and the Poisson distribution were used to investigate the relationship between intersection characteristics and accident counts. We chose the Poisson distribution because it is a common choice for modeling count data and it inherently accounts for the fact that counts cannot be negative.

The first model (Model 1) does not include road visibility percentage, and can be defined as follows:

\begin{align*}
y_i &\sim \text{Poi}(\lambda_i) \\
\hat{\lambda}_i &= \exp\left(\beta_0 + \beta_1 \cdot \text{{traffic}}_i + \beta_2 \cdot \text{{max\_speed}}_i + \beta_3 \cdot \text{{road\_type\_primary}}_i + \beta_4 \cdot \text{{road\_type\_secondary}}_i\right) \\
\end{align*}
\begin{align*}
\text{where:} \\
\text{{traffic}}_i & : \text{The volume of traffic at intersection } i \\
& \text{This is a continuous variable, measured as vehicles passing through.} \\
\text{{max\_speed}}_i & : \text{The maximum speed limit at intersection } i \\
& \text{This is a continuous variable, measured in kilometers per hour.} \\
\text{{road\_type\_primary}}_i & : \text{A binary variable indicating intersection } i \text{ is located on a primary road.} \\
& \text{1 for primary roads, 0 otherwise.} \\
\text{{road\_type\_secondary}}_i & : \text{A binary variable indicating intersection } i \text{ is located on a secondary road.}\\
& \text{1 for secondary roads, 0 otherwise.} \\
\end{align*}

The second model (Model 2) includes the road visibility percentage as an additional predictor:

\begin{align*}
y_i &\sim \text{Poi}(\lambda_i) \\
\hat{\lambda}_i &= \exp\Bigg(\beta_0 + \beta_1 \cdot \text{{road\_visible\_percentage}}_i + \beta_2 \cdot \text{{traffic}}_i + \beta_3 \cdot \text{{max\_speed}}_i \\
&\qquad + \beta_4 \cdot \text{{road\_type\_primary}}_i + \beta_5 \cdot \text{{road\_type\_secondary}}_i\Bigg)
\end{align*}
\begin{align*}
\text{where:} \\
\text{{road\_visible\_percentage}}_i & : \text{The percentage of the view from intersection } i \text{ that is not blocked.} \\
& \text{This is a continuous variable ranging from 0 to 1} \\
\end{align*}

The selection of predictors in our models was guided by prior literature and logical reasoning, aiming to encompass potential factors that might impact accident occurrence at intersections. A unique and novel factor that we introduced in our study is the road visibility percentage. We hypothesized that this factor might significantly influence accident risk, given its direct correlation with the driver's ability to perceive and react to potential hazards. By incorporating these predictors into our models, we aim to achieve a more comprehensive understanding of the factors shaping intersection safety and accident probabilities.

\subsubsection{Model Justification}

The rationale behind our selection of GLMs lies in their adaptability and robustness. GLMs offer versatility in handling a broad spectrum of data distributions and link functions, making them suitable for modeling the count data we have in our study. The Poisson distribution, which inherently accommodates non-negative count data, aligns perfectly with our research requirements, making it a preferred choice for our modeling endeavors.

Furthermore, the Poisson GLM has been widely used in previous traffic safety studies due to its efficacy in capturing the discrete, non-negative, and often skewed distribution of accident counts. To assess the potential impact of road visibility percentage on the model's predictive performance, a comparative analysis was conducted between two models, one that incorporates road visibility percentage as a predictor and another that excludes it.

\section{Results and Discussion}

In this section, we will discuss the experimental setup and the results derived from our analysis. The experiment was performed on the city of London's road accident data, offering a quantitative examination of the relationship between intersection visibility and the occurrence of accidents.

\subsection{Experiment Setup}

\begin{figure}[ht]
\centering
\includegraphics[width=0.7\textwidth]{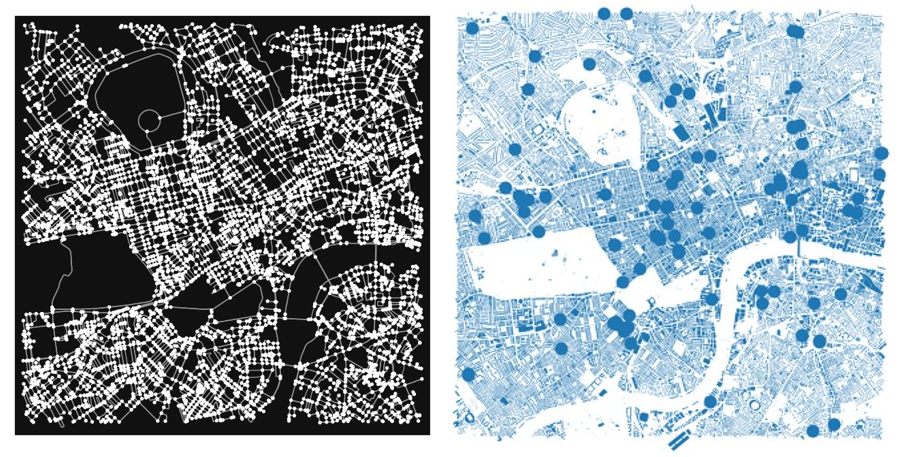}
\caption{The plot of OSM road network (left) and building geographic (right) of London used in this experiment}
\label{fig:OSM}
\end{figure}

The experiment was executed in a controlled environment, concentrating on the city of London, with a designated location point (51.50212, -0.19123) serving as the center. Geographical data within a 3000-meter radius from this point was collected. The OSM road network and building geographic are shown in Figure \ref{fig:OSM}.

The geographical data, encompassing road networks and buildings, was sourced from OpenStreetMap (OSM). The road accident data was retrieved from the UK accident dataset, which offers comprehensive information about the accidents, including their locations, times, and severity levels.

Intersections were identified from the OSM data, with their visibility calculated using the proposed method. This method involves determining the unobstructed view area at each intersection, considering the potential obstructions created by surrounding buildings. Subsequently, accident data were then overlaid onto the road network, linking each accident with its nearest intersection. This setup allowed us to examine the correlation between intersection visibility and accident frequency and identify any significant patterns or trends.

\subsection{Dataset Overview}

Our study utilized a dataset consisting of 935 records, each representing a unique road intersection in London. This dataset was derived after applying several preprocessing steps, including the elimination of intersections with no recorded accidents and merging of intersections that were too close together. These steps were taken to ensure the robustness and reliability of our analysis. The refined dataset includes various attributes of road intersections and associated accident statistics. The variables considered in our analysis include:

\begin{table}[ht]
\centering
\caption{Summary Statistics for the Dataset}
\begin{tabular}{lccc}
\hline
\textbf{Variable} & \textbf{Range/Categories} & \textbf{Mean} & \textbf{SD} \\
\hline
\hline
accidents\_2010\_2021 & 2-114 & 13.16 & 16.33 \\
visible\_percentage & 0.35-0.92 & 0.64 & 0.09 \\
road\_type & secondary, primary & -& - \\
& (66.60\%),  (33.40\%) &  & \\
max\_speed & 20-30 & 20.19 & 1.35 \\
traffic & 0-5269109361 & 9252944 & 174662408 \\
\hline
\end{tabular}
\label{table:summary_stats}
\end{table}

The dataset, presented in Table \ref{table:summary_stats}, includes five variables for each road section: 'road visible percentage', 'traffic', 'max speed', 'road type primary', and 'road type secondary'. These variables provide a comprehensive view of road conditions and accident history in the area under study.

The 'road visible percentage' variable, which measures the proportion of road visible at an intersection, was found to be a significant predictor of accident rates in our models. The 'traffic' variable, reflecting traffic volume, had a range of values, mirroring the variability in traffic volume across different road sections. The 'max speed' variable indicates the maximum speed limit on each road section. The 'road type primary' and 'road type secondary' are binary variables indicating whether a road section is of primary or secondary type. These variables were found to significantly impact the number of accidents in our models, with primary roads associated with a higher risk of accidents.

\subsection{Regression Analysis}

The Generalized Linear Model (GLM) was employed to examine the relationship between road visibility and accident occurrence. Two models were compared: one considering visibility (Model 2) and one excluding it (Model 1).

\begin{table}[ht]
\centering
\caption{Regression Report for Model 1 and Model 2}
\begin{tabular}{lll}
\hline
\textbf{Metric} & \textbf{GLM Poisson Model} & \textbf{GLM Poisson Model with Visibility} \\
\hline
\hline
Observations & 967 & 967 \\
Residuals & 963 & 962 \\
Link Function & Log & Log \\
Method & IRLS & IRLS \\
Log-Likelihood & -7699.5 & -7537.1 \\
Deviance & 11655 & 11330 \\
Pearson chi2 & 1.75e+04 & 1.64e+04 \\
Iterations & 5 & 5 \\
Pseudo R-squ. (CS) & 0.9046 & 0.9318 \\
\hline
\end{tabular}
\label{table:regression_results_1}
\end{table}

\begin{table}[ht]
\centering
\caption{Regression Results for Model 1 and Model 2}
\begin{tabular}{lccccc}
\hline
& \multicolumn{2}{c}{\textbf{Model 1}} & \multicolumn{2}{c}{\textbf{Model 2}} \\
\cline{2-3} \cline{4-5}
& Coef. & P$>|$z$|$ & Coef. & P$>|$z$|$ \\
\hline
\hline
const & 0.4476 & 0.000 & -0.2075 & 0.003 \\
visible\_percentage & & & 1.7450 & 0.000 \\
traffic & -1.731e-11 & 0.780 & 4.388e-11 & 0.467 \\
max\_speed & 0.0967 & 0.000 & 0.0886 & 0.000 \\
road\_type\_primary & 0.6150 & 0.000 & 0.2543 & 0.000 \\
road\_type\_secondary & -0.1674 & 0.000 & -0.4617 & 0.000 \\
\hline
\end{tabular}
\label{table:regression_results_2}
\end{table}

The regression results are presented in Tables \ref{table:regression_results_1}, \ref{table:regression_results_2}, and \ref{table:aic_bic_comparison}. As shown, the inclusion of the 'visible percentage' variable in Model 2 improved the model fit and explanatory power significantly.

As shown in Table \ref{table:regression_results_1}, Model 2 outperforms Model 1 based on several metrics. This is evident from Model 2's higher Log-Likelihood, lower Deviance and Pearson chi-squared, and higher Pseudo R-squared (Cragg-Uhler), suggesting a better fit to the data.

In Table \ref{table:regression_results_2}, the coefficients and significance levels of the variables in both models are presented. Most notably, the 'visible percentage' variable, included only in Model 2, is highly significant and has a large coefficient, indicating a strong positive relationship with the number of accidents.

Table \ref{table:aic_bic_comparison} presents a comparison of the Akaike Information Criterion (AIC) and the Bayesian Information Criterion (BIC) for both models. Lower values of AIC and BIC are preferable as they indicate a better balance of fit and complexity. Model 2 has lower AIC and BIC values than Model 1, again suggesting that it is the more optimal model.

These findings underline the importance of considering visibility at intersections when predicting accident risks. The 'visible percentage' variable has proven to be a significant predictor in our model, contributing to an improved understanding of the factors that influence road accidents.

\begin{table}[ht]
\centering
\caption{AIC and BIC comparison for Model 1 and Model 2}
\begin{tabular}{lcc}
\hline
\textbf{Model} & \textbf{AIC} & \textbf{BIC} \\
\hline
\hline
Model 1 & 15406.98 & 5035.25 \\
Model 2 & 15084.18 & 4717.32 \\
\hline
\end{tabular}
\label{table:aic_bic_comparison}
\end{table}

\subsection{Investigation on Model Performance and Findings}

Our study utilized Generalized Linear Models (GLMs) with a Poisson distribution to investigate the influence of various road characteristics on the number of accidents. This model was suitable due to the count nature of our response variable (the number of accidents), which aligns with a Poisson distribution.

The regression analyses yielded several notable insights:

\begin{itemize}
\item Road type significantly impacts the number of accidents. Specifically, intersections located on 'primary' roads were associated with an increase in the log of expected count of accidents by 0.615 (p<0.001), while those on 'secondary' roads displayed a decrease of -0.167 (p<0.001). These effects were statistically significant, suggesting that road type is a critical predictor of accident occurrence.

\item Our findings indicate that the 'view percentage' at an intersection - a measure of visibility - plays a significant role in predicting accident risks. An increase in the view percentage by 1 unit was associated with an increase in the log of expected count of accidents by 1.745 (p<0.001). However, the underlying reasons for this relationship are not fully understood. Further research, preferably utilizing more detailed and comprehensive data, is needed to investigate the underlying mechanisms that link intersection visibility to accident occurrence.

\item Traffic volume and maximum speed limit also considerably influenced accident occurrence. A unit increase in traffic volume and maximum speed limit was associated with a change in the log of expected count of accidents by 4.388e-11 (p=0.467) and 0.0886 (p<0.001) respectively. These findings align with existing literature in the field.

\item With regard to model fit, the Pseudo R-squared values suggested that our models explained a significant proportion of the variability in accident counts, indicating a good fit to the data. The Pseudo R-squared value for the model including visibility was 0.9318, higher than the model without visibility (0.9046), suggesting the added value of the visibility measure in our predictive model.
\end{itemize}

These results offer valuable insights into the factors affecting road accidents at intersections. However, further research is warranted to corroborate these relationships and delve deeper into the underlying mechanisms. As such, our findings serve as a useful starting point for future studies aimed at understanding these complex relationships and formulating effective road safety measures.

\subsection{Testing on Manchester City}

As part of our study, we expanded our analysis to another city, Manchester, to verify the robustness of our models and findings. We applied the same preprocessing and feature engineering steps to the Manchester data, and then fitted the Poisson and Negative Binomial regression models.

\begin{figure}[ht]
\centering
\includegraphics[width=0.6\textwidth]{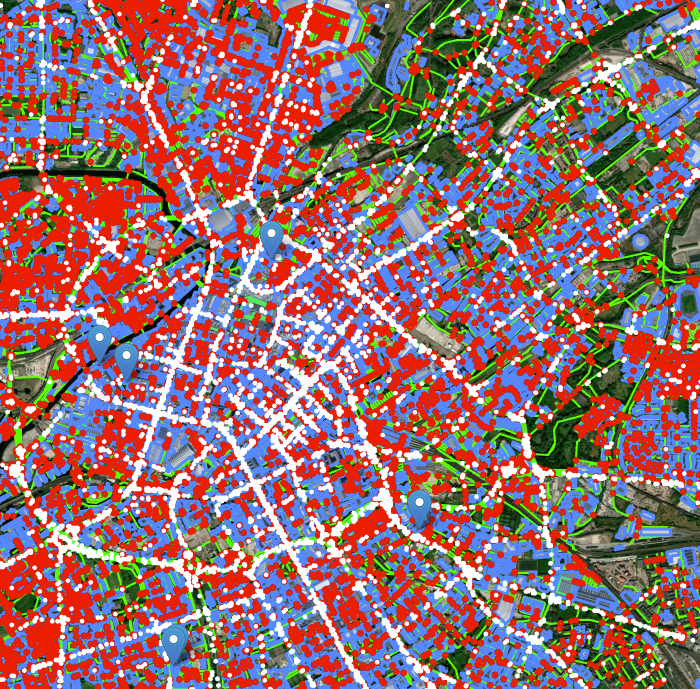}
\caption{Folium map of the Manchester City map used in this experiment.}
\label{fig:occlusion}
\end{figure}

\begin{table}[htbp]
\centering
\caption{Generalized Linear Model Regression Results on Manchester City}
\begin{tabular}{ll}
\hline
\textbf{Metric} & \textbf{Feature Values}  \\
\hline
No. Observations & 243 \\
Df Residuals & 238 \\
Link Function & Log \\
Method & IRLS \\
Log-Likelihood & -1938.5 \\
Deviance & 2836.7 \\
Pearson chi2 & 4.24e+03 \\
No. Iterations & 6 \\
Pseudo R-squ. (CS) & 0.8664 \\
\hline
\end{tabular}
\label{table:manchester_model_results}
\end{table}

The dataset, regression models, and results for Manchester city are presented in Figure \ref{fig:occlusion} and Table \ref{table:manchester_model_results}. The findings of this study, which align closely with those obtained from the London data, reinforce the validity and generalizability of our models. The results suggest that the key road characteristics - road type, road visible percentage, traffic volume, and maximum speed limit - consistently influence the number of accidents at intersections, regardless of the city. This successful replication of results across different cities underscores the robustness of our approach. Future studies could further validate these findings by applying our models to data from a broader range of cities.

\section{Conclusions and Future Work}

Our research provides innovative insights into the impact of intersection visibility and road characteristics on road accidents in the UK. Through the introduction of our unique 'view percentage' measure to quantify intersection visibility, we were able to establish a substantial relationship between visibility and accident rates. Our models, validated using AIC and BIC, affirmed the importance of intersection visibility in predicting accident occurrences. These findings contribute valuable knowledge to the field of road safety and highlight the importance of considering intersection visibility in accident prediction and prevention strategies.

The robustness of our approach is reinforced by consistent findings across datasets from both London and Manchester. Particularly intriguing is the observation of a positive relationship between visibility and accident rates, indicating that higher visibility may potentially be associated with more aggressive driving behaviors.

While our study provides valuable insights, it also unveils avenues for future research in this domain. Notably, a more detailed accident record could shed further light on how the view percentage influences accident occurrence and road risk. Understanding the precise circumstances of each accident, such as the driver's view at the time, could help tease out the underlying mechanisms of this relationship.

Moreover, future studies could broaden the geographic scope of the study to encompass other regions, explore the impact of neighboring intersections on accident risk, and enrich our models with more granular traffic conditions and driver behavior data.

\section{Author Contribution}
The authors confirm contribution to the paper as follows: study conception and design: Hanlin Tian, Yuxiang Feng, Wei Zhou, Anupriya, Mohammed Quddus, Yiannis Demiris, and Panagiotis Angeloudis; data collection: Hanlin Tian, Wei Zhou; analysis and interpretation of results: Hanlin Tian, Anu Priya; draft manuscript preparation: Hanlin Tian, Yuxiang Feng, Wei Zhou. All authors reviewed the results and approved the final version of the manuscript.

\newpage

\bibliographystyle{trb}
\bibliography{trb_template}
\end{document}